\documentclass[english,format=sigconf, screen=true]{acmart}
\usepackage[T1]{fontenc}
\usepackage[latin9]{inputenc}
\PassOptionsToPackage{ruled}{algorithm2e}
\usepackage{booktabs}
\usepackage{algorithm2e}
\usepackage{amsbsy}
\usepackage{amssymb}
\usepackage{graphicx}

\makeatletter
\copyrightyear{2020}
\acmYear{2020}
\acmBooktitle{KDD Workshop on Machine Learning in Finance (MLF '20), August 24, 2020, Virtual Event, USA}
\acmDOI{}

\providecommand{\tabularnewline}{\\}

\renewcommand\footnotetextcopyrightpermission[1]{} 
\usepackage{dsfont}
\SetAlFnt{\small}
\SetAlCapFnt{\small}
\SetAlCapNameFnt{\small}
\SetAlCapHSkip{0pt}
\IncMargin{-\parindent}

\makeatother

\usepackage{babel}
\begin{document}
\fancyhead{}
\title{Towards Earnings Call and Stock Price Movement}
\author{Zhiqiang Ma}
\affiliation{\institution{S\&P Global}\city{New York}\state{NY}\country{USA}}
\email{zhiqiang.ma}
\email{@spglobal.com}
\author{Grace Bang}
\affiliation{\institution{S\&P Global}\city{New York}\state{NY}\country{USA}}
\email{grace.bang}
\email{@spglobal.com}
\author{Chong Wang}
\affiliation{\institution{S\&P Global}\city{New York}\state{NY}\country{USA}}
\email{chong.wang}
\email{@spglobal.com}
\author{Xiaomo Liu}
\affiliation{\institution{S\&P Global}\city{New York}\state{NY}\country{USA}}
\email{xiaomo.liu}
\email{@spglobal.com}
\begin{abstract}
Earnings calls are hosted by management of public companies to discuss
the company\textquoteright s financial performance with analysts and
investors. Information disclosed during an earnings call is an essential
source of data for analysts and investors to make investment decisions.
Thus, we leverage earnings call transcripts to predict future stock
price dynamics. We propose to model the language in transcripts using
a deep learning framework, where an attention mechanism is applied
to encode the text data into vectors for the discriminative network
classifier to predict stock price movements. Our empirical experiments
show that the proposed model is superior to the traditional machine
learning baselines and earnings call information can boost the stock
price prediction performance.
\end{abstract}



\maketitle

\section{Introduction\label{sec:Introduction}}

With \$74 trillion in assets under management in the US alone\footnote{https://www.bloomberg.com/graphics/2019-asset-management-in-decline},
understanding the mechanism of stock market movements is of great
interest to financial analysts and researchers. As such, there has
been significant research in modeling stock market movements using
statistical and, more recently, machine learning models in the past
few decades. However, it may not be sensible to directly predict future
stock prices given the possibility that they follow a random walk
\citep{Malkiel1999}. Thus researchers have proposed to predict the
directional movements of stocks and their volatility levels \citep{Theil2019,Hu2018,Xu2018}.
In this study, we explore company's earnings call transcripts data
and investigate using the information embedded in earnings call transcripts
to address the task of predicting the movements of stocks by leveraging
the recent advancements in natural language processing (NLP).

Stock markets demonstrate notably higher levels of volatility, trading
volume, and spreads prior to earnings announcements given the uncertainty
in company performance \citep{Donders2000}. Such movements can be
costly to the investors as they can result in higher trading fees,
missed buying opportunities, or overall position losses. Thus, the
ability to accurately identify directional movements in stock prices
and hold positions accordingly based on earnings releases can be hugely
beneficial to investors by potentially minimizing their losses and
generating higher returns on invested assets. 

Stock market prices are driven by a number of factors including news,
market sentiment, and company financial performance. Predicting stock
price movements based on market sentiment from the news and social
media have been studied previously \citep{Hu2018,Ding2015,Xu2018}.
However, earnings calls, which occur when companies report on and
explain their financial results, have not been extensively studied
for predicting stock price movements.

Earnings call are conference calls hosted by the companies and occur
between the senior executives of publicly traded companies and call
participants such as investors and equity analysts. Generally, the
earnings calls are comprised of two components: 1) Presentation of
recent financial performance by senior company executives and 2) Question
and Answer (Q\&A) session between company management and market participants.
Earnings calls are comprised of tremendous insights regarding current
operations and outlook of companies, which could affect confidence
and attitude of investors towards companies and therefore result in
stock price movements. The first part of the earnings call -- Presentation
-- is typically scripted and rehearsed, particularly in the face
of bad news. However, the question and answer portion of the call
incorporates unscripted and dynamic interactions between the market
participants and management thus allowing for a more authentic assessment
of a company. Thus, we focus on the Answer section in this work and
discuss our findings regarding Presentation data in Section \ref{sec:Discussion-and-Conclusions}.

In this paper, we propose a deep learning network to predict the stock
price movement, in which sentences from the Answer section of a transcript
are represented as vectors by aggregating word embeddings and an attention
mechanism is used to capture their contributions to predictions. Discrete
industry categories of companies are also considered in the work by
encoding them into learnable vector representations. We compare the
proposed method with several classical machine learning algorithms
to assess its effectiveness. We review several related work and present
our researching and findings in the reset of this paper.

\section{Related Work}

Stock price movement predictions have traditionally been considered
a time series prediction problem \citep{Xu2018}. Existing approaches
tackle this problem by discovering trading patterns in the historical
market data to predict future movements. Statisticians usually use
time series analysis techniques like exponential smoothing, autoregressive
(AR), and autoregressive integrated moving average (ARIMA) to predict
prices or price movements. Computer science researchers have also
showed great interest in this topic and have applied machine learning
prediction models \citep{Sheta2015,Patel2014} to solving this task.
Recurrent neural networks (RNN) and especially its variants such as
LSTM \citep{Hochreiter1997}, which were developed to process sequential
signals, have been widely adopted to model time series stock data
\citep{Nelson2017,Cheng2018}. In contrast to these statistical methods,
RNN is not subject to the stationarity requirement on the stock time
series data and is able to capture the dependency of stock prices
at different time instances.

Another important research branch on this topic concentrates on leveraging
external information outside of market data, e.g., events, news, macroeconomic
environment, business operations, and geopolitical status, as drivers
of stock price movements. For example, Equifax\textquoteright s stock
price plummeted more than 15 percent immediately following news reports
that it had suffered a massive data breach scandal. \citep{Ding2015,Deng2019}
proposed to extract structured events from news and then use deep
neural networks to model the impact of the events on the stock movement.
Hu et al. \citep{Hu2018} developed a hierarchical attention based
neural network -- HAN -- studying the dependency and influence of
the recent online news on stock markets. As social media began reporting
breaking news, researchers found social media posts can serve as input
along with historical stock data \citep{Xu2018,Si2013}. Financial
filings (10-K) summarizing companies' business performance contain
sentiment signals from management, which can be used to forecast stock
return volatility \citep{Wang2013}. Bag-of-words features, TFIDF
and LOG1P, were adopted in their work. Researchers name this type
of prediction fundamental analysis, and since our work adopted earnings
call as the input, it falls in this category as well.

As mentioned in the Section \ref{sec:Introduction}, earnings call
transcripts have unique properties and provide crucial information
of companies. There is tremendous potential for exploration of this
dataset as limited previous work has studied earnings call transcripts
in stock return volatility prediction to evaluate companies' financial
risk \citep{Wang2014,Theil2019}. In \citep{Theil2019}, Theil et
al. introduced a neural network PRoFET to predict the stock return
volatility, where it considers textual features from earnings call
transcripts and financial features including past volatility, market
volatility, book-to-market, etc. To create the textual feature, each
section (presentation, questions, and answers) is represented as a
vector by applying a Bi-LSTM with attention mechanism \citep{Bahdanau2015}
on the tokens. Financial features pass through a deep feedforward
network to calculate the financial vector. The final prediction is
returned by summing these two vectors and feeding to another hidden
layer.

In NLP tasks, word representation is always a critical component.
Pre-trained word vectors and embeddings have been widely adopted in
various state-of-the-art NLP architectures and achieved great success.
The work like word2vec \citep{Mikolov2013} and GloVe \citep{Pennington2014}
represents words as high dimensional real-valued vectors, and their vector
arithmetic operations can reflect the semantic relationship of the
words. In this work, we adopted the pre-trained GloVe embeddings to
save computing time.

\section{Problem Statement}

Assuming that there is a set of stocks $\boldsymbol{\Theta}=\{S_{1},S_{2},\cdots,S_{n}\}$
of $n$ public companies. For a stock $S_{c}$, there exists a series
of earnings call transcript $\boldsymbol{\Gamma}_{c}=\{\boldsymbol{T}_{d_{1}},\boldsymbol{T}_{d_{2}},\cdots,\boldsymbol{T}_{d_{m}}\}$,
which were held on dates $d_{1},d_{2},\cdots,d_{m}$ respectively.
The goal is to predict the movement of the stock $S_{c}$ on date
$d_{i}+1$ given the earnings call $\boldsymbol{T}_{d_{i}}$ occurred
on date $d_{i}$. The movement $y$ is a binary value, $0$ (down)
or $1$ (up). The stock price in the market moves constantly in a
trading day. To formally define $y$, here we adopt the closing price,
i.e. $y=\mathds{1}(p_{d_i+1}>p_{d_i})$, where $p_{d_{i}}$ and $p_{d_{i}+1}$
are the closing prices on date $d_{i}$ and $d_{i}+1$.

We aim to learn a prediction function $f$, which takes features $\boldsymbol{E}$
extracted from an earnings call transcript $\boldsymbol{T}$ and industry
categorization $\boldsymbol{I}$ of the company as input, to predict
the stock price movement $y$ of the day after the earnings call.

\section{Model Overview}

To solve the problem, we utilize two features to build the prediction
function: 1) Answer section textual feature and 2) company industry
type feature. In this section, we firstly propose a deep neural network
structure designed to represent the textual feature. Hereafter, we
introduce the industry type embedding. The final prediction is generated
via a discriminative network by feeding in the combined features.

\subsection{Earnings Call Representation}

A Q\&A section consists of multiple rounds of communications between
market participants and company management executives. We only use
Answer sections from managements with the assumption that the answers
are a more realistic representation of the feedback interested by
investors. In the case where a response provided by managements does
not answer a specific question, market participants typically follow
up with clarifying questions to which they then receive required answers.

\subsubsection*{Sentence Embedding:}

Given an earnings call transcript $T$, we extract the answer sequence
$\boldsymbol{A}=[l_{1},l_{2},\cdots,l_{N}]$ and $\boldsymbol{A}\in\boldsymbol{T}$,
$l_{i}$ denoting a sentence that comes from splitting the Answer
section. We treat one sentence as a feature atom, and transform each
sentence to a dense vector.  To achieve that, we process each token
$o$ of a sentence $l$ to a distributed representation vector $\boldsymbol{e}_{o}$
by leveraging a pre-trained embedding layer. The sentence vector $\boldsymbol{v}_{l}$
is constructed by concatenating two vectors obtained from average
pooling and max pooling the token vectors across all the tokens of
the sentence. To reduce computing complexity, we didn't allow the
word embedding layer to be trainable or fine-tuned. Another popular
approach to representing sentences is to employ RNN to encode a whole
sentence to a hidden state vector from the last recurrent unit \citep{Theil2019}.
Sentence encoders \citep{Conneau2017,Cer2018} may be used here too.
We leave them for the future exploration.

\subsubsection*{Sentence Attention:}

\begin{figure}
\includegraphics[clip,scale=0.8]{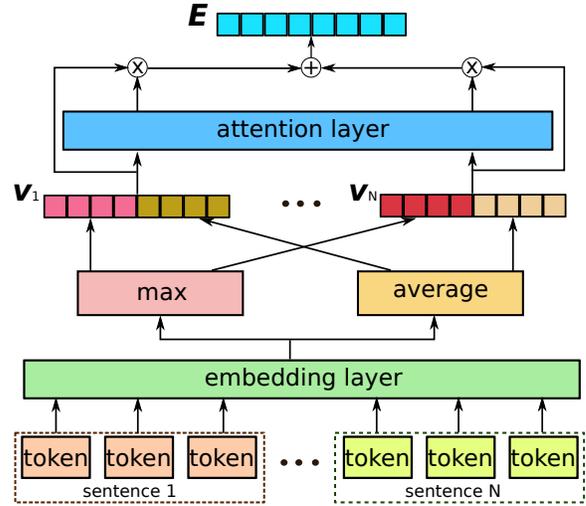}\caption{\label{fig:earnings_emb}Neural network structure for learning textual
feature vectors. The input is tokens from sentences of an Answer section,
and the output $\boldsymbol{E}$ is a vector representation of the
input.}
\end{figure}

Undoubtedly, some sentences convey more information while others do
not for the task of predicting stock price movements. We leverage
the idea of the attention mechanism introduced in the machine translation
domain \citep{Bahdanau2015} to learn the weights of the sentences,
where the weights quantify the contributions of the sentences to the
final outcome. Given an answer sequence $\boldsymbol{A}$ consisting
of $N$ sentences and sentences transformed to embedding vector $\boldsymbol{v}$s,
the attention weights $\boldsymbol{\alpha}\in\mathbb{R}^{1\times N}$
are defined as normalized scores over all the sentences by a softmax
function as shown below,
\begin{equation}
\begin{array}{cc}
\alpha_{l}=\mathrm{softmax}(\mathrm{score}(\boldsymbol{v}_{l})),\\
\mathrm{score}(\boldsymbol{v}_{l})=\boldsymbol{u}^{T}\boldsymbol{v}_{l}+b,
\end{array}\label{eq:attention_1}
\end{equation}
where $\boldsymbol{u}$ is a learnable vector parameter and $b$ is
a learnable bias parameter. The score function may be replaced with
others depending on the specific task. Refer to \citep{Bahdanau2015,Luong2015,Vaswani2017}
for other score function options. By aggregating the sentence vectors
weighted on the attention parameter, the earnings call answer sequence
can be transformed to 
\begin{equation}
\boldsymbol{E}=\sum_{l}^{N}\alpha_{l}\boldsymbol{v}_{l}.\label{eq:attention_2}
\end{equation}
Figure \ref{fig:earnings_emb} demonstrates our network structure
introduced above.

\subsection{Industry Embedding}

Company stock prices usually follow the trend of the industry sector
in which it belongs. The sector category and company sector definition
vary in terms of standards. We select the Global Industry Classification
Standard (GICS) definition in our study. GICS consists of 11 industry
sector categories, such as `energy', `financials', and `health care'.
The industry sector is a categorical indicator.  In machine learning,
categorical data are usually transformed by one-hot encoding or ordinal
encoding, while we create an embedding layer to transform the categorical
values into vector presentation $\boldsymbol{I}$, which is learnable
during the network training phase. 

\subsection{Discriminative Network Structure}

\begin{figure}
\includegraphics[clip,scale=0.8]{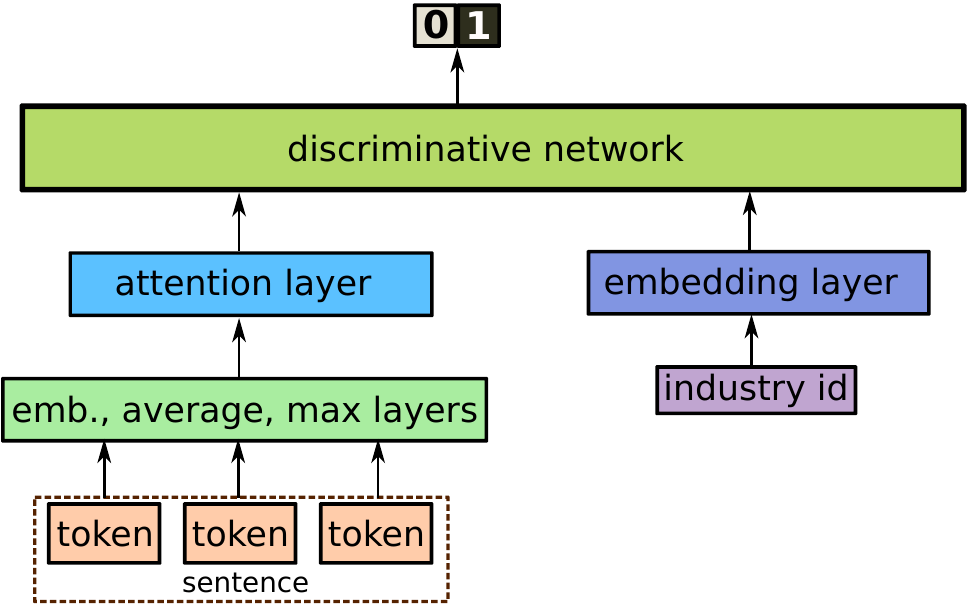}\caption{Proposed neural network structure. The input to the discriminative
network is a concatenated vector of a textual feature vector and an
industry embedding vector. \label{fig:dis-network}}
\end{figure}

With the feature representations $\boldsymbol{E}$ and $\boldsymbol{I}$
built above as input, the final binary classification result is computed
by a discriminative network. The feed forward discriminative network
consists of multiple hidden layers --- batch normalization layer
\citep{Ioffe2015}, dropout layer \citep{Srivastava2014}, ReLU activation
layer \citep{Nair2010}, and linear layer. Figure \ref{fig:dis-network}
illustrates the complete neural network structure including the discriminative
network.

\section{Experiment Studies}

\subsection{Data}

\begin{figure}
\includegraphics[bb=15bp 30bp 640bp 520bp,clip,scale=0.25]{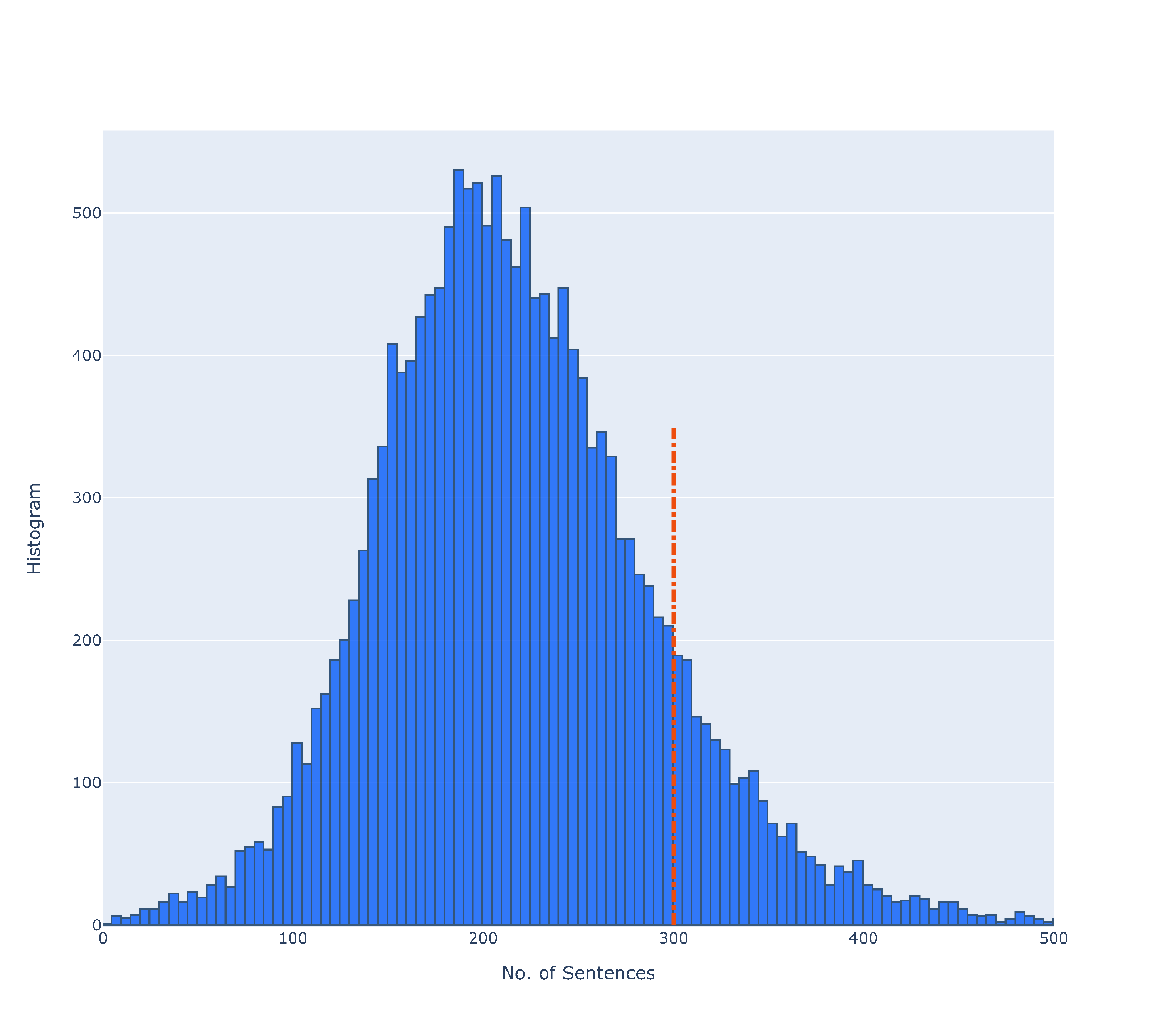}\caption{Histogram of the number of sentences in Answer sections. The dashed
line in red at 300 at $\mathtt{x-axis}$ indicates the cut-off on
the length of the Answer section. \label{fig:Histogram-of-sentences}}
\end{figure}

We perform experiments by using earnings calls transcripts of S\&P
500 companies. We collected 17025 earnings call transcripts over 485
companies\footnote{S\&P 500 index composes 505 stocks from 500 companies. Due to merger
and acquisition, ticker changing, shortness of available transcripts,
and etc. reasons, 15 companies were not included in the data.} from S\&P Global Market Intelligence TRANSCRIPTS database. The temporal
span of the data is between 2009 and 2019, and the temporal spans
for a few companies might be shorter because the companies were added
to S\&P 500 later than 2009. On average, each company has around 35
transcripts. Every transcript in the TRANSCRIPTS database has been
segmented into components in terms of types of the components, such
as `Presentation Operator Message', `Presentation Section', `Question',
and `Answer'. We select the `Answer' components and employed NLTK
sentence tokenizer to split sections to sentences. Figure \ref{fig:Histogram-of-sentences}
shows the distribution of the number of sentences in `Answer' components
in the dataset. Table \ref{tab:transcript_data_stat} shows more statistics
of the dataset in terms of sentences and tokens (stop words excluded).
As to download the corresponding historical stock data to get the
stock price movements $\boldsymbol{y}$, we map company names to their
stock tickers and employed Python \texttt{pandas\_datareader} package
with the source set to \texttt{yahoo}.

\begin{table}
\begin{tabular}{ccc}
\toprule 
Answer Section & No. of Sentences & No. of Tokens\tabularnewline
\midrule 
Total & 3.6 \texttt{M} & 21.7 \texttt{M}\tabularnewline
Average & 2145 & 1457\tabularnewline
\bottomrule
\end{tabular}\caption{Statistics of the raw text data (without stop words).\label{tab:transcript_data_stat}}
\end{table}

\subsection{Experiments}

\subsubsection{Model Training Settings:}

For the experiments, we hold out the most recent five earnings call
transcripts from each company as the testing dataset (2425 observations
in total), and everything else is used as the training and validation
dataset. Please note that companies have their earnings call conferences
on different dates for every reporting quarter, and it is not reasonable
to set a cutoff date to split the dataset.

We split Answer section to sentences and then tokenize sentences to
tokens. When transforming tokens to embedding vectors, a vocabulary
is constructed, where stop words are ignored and tokens with total
frequency less than four times would be disregarded as well. Tokens
are transformed to vectors by applying pre-trained GloVe (embedding
dimension $=300$). As Figure \ref{fig:Histogram-of-sentences} describes
with the red cut-off line, when learning the attention scores for
the sentences, we set the dimension of $\boldsymbol{E}$ of each transcript
to $N=300$, i.e., transcripts composing more than 300 sentences would
be truncated or padded if less than 300 in our implementation. Transcripts
with their Answer section lengths shorter than ten sentences are ignored.
The model is implemented in Pytorch and experimented on a Nvidia V100
GPU server.

\subsubsection{Baselines:}

In order to assess the performance of our model, we compare its performance
with two baseline models below:
\begin{itemize}
\item \textit{Mean Reversion (MR)}: MR is a simple trading strategy, which
assumes that stock prices would tend to revert toward their moving
averages when deviating from them. We calculate 60-day moving average
in the experiment.
\item \textit{XGBoost}: XGBoost has achieved great success in solving various
classification problems in practice. To transform the text data into
numeric format, we adopted two feature engineering techniques, \texttt{TFIDF}
and \texttt{LOG1P} defined as below \citep{Kogan2009}:
\begin{itemize}
\item $\textrm{TFIDF}=\textrm{TF}(o,\boldsymbol{A})\times\textrm{IDF}(o,\boldsymbol{A})=\textrm{TC}(o,\boldsymbol{A})\times\textrm{IDF}(o,\boldsymbol{A})$
\item $\textrm{LOG1P}=\textrm{LOG}(1+\textrm{TC}(o,\boldsymbol{A}))$
\end{itemize}
where $\textrm{TC}(o,\boldsymbol{A}))$ is the count of the token
$o$ in earnings call Answer section $\boldsymbol{A}$ and $\textrm{IDF}(o,\boldsymbol{A})=\textrm{log}(|\boldsymbol{\Gamma}|/|\boldsymbol{A}\in\boldsymbol{\Gamma},o\in\boldsymbol{A}|)$.
\end{itemize}

\subsubsection{Results:}

In the experiments, we predict the stock price movement of the companies
on the day just after their earnings call being released. Table \ref{tab:exp_acc_mcc}
compares the performance of our proposed model and the baseline models.
When compare the models, we adopt two evaluation metrics, accuracy
and Matthews Correlation Coefficient (MCC), which are also used in
the previous work \citep{Xu2018,Ding2015}. The definition of MCC
is as follows, given true positive (tp), true negative (tn), false
positive (fp), and false negative (fn) from the prediction output:

\[
\textrm{MCC}=\frac{\textrm{tp}\cdot\textrm{tn}-\textrm{fp}\cdot\textrm{fn}}{\sqrt{(\textrm{tp}+\textrm{fp})(\textrm{tp}+\textrm{fn})(\textrm{tn}+\textrm{fp})(\textrm{tn}+\textrm{fn})}}.
\]
 The value of MCC is between $-1$ and $1$, where $1$ stands for
complete match between predictions and ground truths while $-1$ means
predictions and ground truths are entirely opposite. It can be seen
that our model outperforms the baseline models by more than 1\% in
the prediction accuracy and doubling the MCC measure.

\begin{figure}
\includegraphics[bb=8bp 12bp 600bp 400bp,clip,scale=0.4]{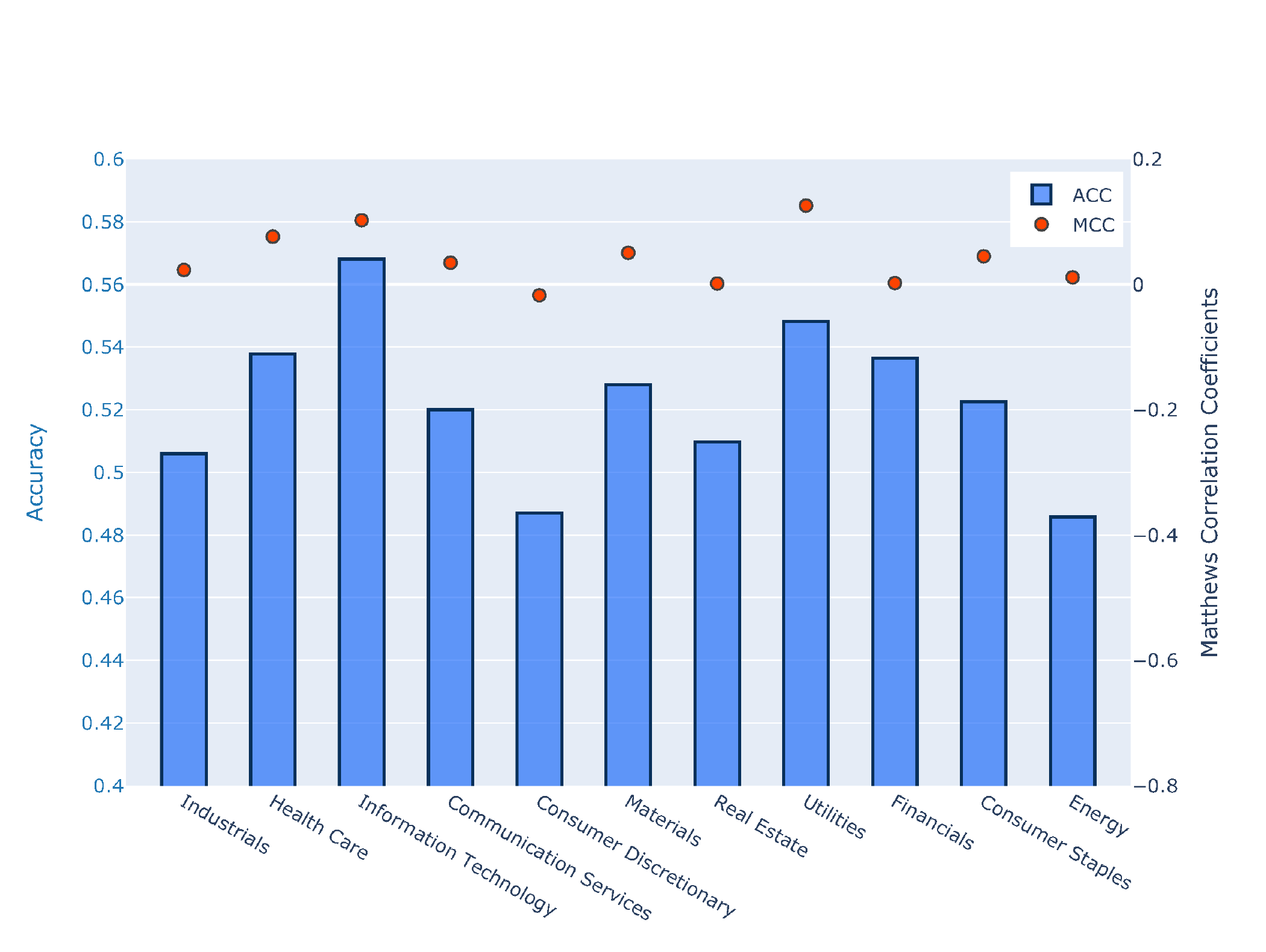}\caption{Model performance in terms of industry sectors. The bars represent
the accuracy measure and the dots indicate the MCC measure.\label{fig:ind_perf}}
\end{figure}

Figure \ref{fig:ind_perf} is the accuracy and MCC measures in terms
of the 11 industry sectors. We can observe that the model performance
varies with respect to industries, the highest accuracy (56.8\%) occurring
in the information technology sector and the lowest accuracy (48.5\%)
in the energy sector. This result is roughly consistent with our common
perception on the stock market -- generally stock price movements
of high-tech companies are driven by bearish or bullish signals from
various sources, while for energy companies their stock performance
heavily relies on the crude oil price and macroeconomic factors rather
than external news and information.

\begin{table}
\caption{\label{tab:exp_acc_mcc}Model performance summary.}

\begin{tabular}{ccc}
\toprule 
 & \multicolumn{1}{c}{Accuracy (\%)} & \multicolumn{1}{c}{MCC}\tabularnewline
\midrule 
MR & 50.80 & 0.0202\tabularnewline
XGBoost (Log1P) & 50.89 & 0.0013\tabularnewline
XGBoost (TFIDF) & 51.25 & 0.0154\tabularnewline
Our Model & \textbf{52.45} & \textbf{0.0445}\tabularnewline
\bottomrule
\end{tabular}
\end{table}

\section{Discussions and Conclusions\label{sec:Discussion-and-Conclusions}}

Undoubtedly, stock price movement prediction is a very challenging
task. Through our experimentation in this study, we confirm that earnings
call transcripts have certain predictive power for future stock price
movements. Thus, the inclusion of this dataset in stock price prediction
analysis can have predictive impact in the development of such systems
for in practice use of stock investment risk analysis. Additionally,
we note two other aspects, which are worth more investigation in the
future:
\begin{itemize}
\item In Section \ref{sec:Introduction}, we mention that only Answer sections
are included in the model with the management Presentation sections
excluded. Our decision on that, besides the heuristics reason mentioned
in Section \ref{sec:Introduction}, is that the model does not improve
by including the Presentation data. The sole consideration of the
Presentation text also did not improve the model performance results.
Interestingly, this observation is not consistent with the conclusion
made by Theil et al. \citep{Theil2019}, where the Presentation data
yields better results in their ablation study for predicting stock
volatility, despite the different prediction targets in their work
and ours. Future work will be conducted to justify the observation.
\item In addition to the fundamental analysis like our work, features originating
from technical analysis on the historical stock price data are able
to be absorbed into the forecast model. For example, the historical
stock time series data can be encoded into another feature vectors
by RNN models, which are further used to build global vectors along
with the features from the fundamental analysis.
\end{itemize}
To summarize, we propose leveraging textual information from Answer
sections of earnings call transcripts to predict movements of stock
prices. To create textual features from transcripts, tokens are transformed
into embedding vectors and sentence vectors are built by max pooling
and average pooling over the word vectors. An earnings call Answer
section is represented as a vector by aggregating its sentence vectors
through the attention mechanism. The final prediction is made by a
discriminative network which takes the textual feature vectors and
learned industry embedding vectors as input. The experiments show
that the proposed deep learning model outperforms the classical baseline
models, and also prove that the information conveyed in earning calls
correlates with stock price movements and therefore can be used in
relevant forecasting tasks.

\bibliographystyle{ACM-Reference-Format}
\bibliography{reference}

\end{document}